\documentclass{mem}
\usepackage{natbib}\usepackage{txfonts}\usepackage{balance}
\usepackage{flushend}
\usepackage{graphicx}
\usepackage[breaklinks,pdftex]{hyperref}
\usepackage{colortbl}
\usepackage{makecell}
\usepackage{graphics,epsfig,amssymb,color,times}
\usepackage{arydshln}
\usepackage{titlesec}
\usepackage[table,xcdraw]{xcolor}
\idline{95}{27}
\begin{document}
\def\teff{$T\rm_{eff }$}
\def\kms{$\mathrm {km s}^{-1}$}

\title{
Unobscured radio-quiet Active Galactic Nuclei under the eyes of {\it IXPE} 
}

\author{
V. E. \,Gianolli\inst{1,2} 
\and S.\, Bianchi \inst{2}
\and P-O. \, Petrucci \inst{1}
\and A. \, Marinucci \inst{3}
\and A. \, Ingram \inst{4}
\and D. \, Tagliacozzo \inst{2}
\and D. E.\, Kim \inst{5,6,7}
\and F. \, Marin \inst{8}
\and G. \, Matt \inst{2}
\and P. \, Soffitta \inst{5}
\and F. \, Tombesi \inst{7,9,10}
}

\institute{
Université Grenoble Alpes, CNRS, IPAG, 38000 Grenoble, France 
\and
Dipartimento di Matematica e Fisica, Università degli Studi Roma Tre, Via della Vasca Navale 84, 00146 Roma, Italy \\ \email{vittoria.gianolli@univ-grenoble-alpes.fr}
\and ASI – Agenzia Spaziale Italiana, Via del Politecnico snc, I-00133 Roma, Italy
\and School of Mathematics, Statistics, and Physics, Newcastle University, Newcastle upon Tyne NE1 7RU, UK
\and INAF Istituto di Astrofisica e Planetologia Spaziali, Via del Fosso del Cavaliere 100, I-00133 Roma, Italy\\
The remaining affiliations can be found at the end of the paper.
}

\authorrunning{Gianolli et al.}

\titlerunning{radio quiet AGN with {\it {\it IXPE}}}

\date{Received: 1 December 2023; Accepted: 15 January 2024}

\abstract{
We present the results of the X-ray polarimetric analyses performed on unobscured radio-quiet Active Galactic Nuclei (AGN) with the Imaging X-ray Polarimetry Explorer ({\it IXPE}), with simultaneous XMM-{\it Newton} and {\it NuSTAR} data. The synergy of these instruments is crucial to constrain the X-ray corona physical properties and assess its geometry. In the first two years of operation, three AGN have been observed: significant polarization was detected for NGC~4151 (4.9$\pm$1.1 per cent) and IC~4329A (albeit with less confidence, 3.3$\pm$1.1 per cent), with polarization angles aligned with their radio jets, while only an upper limit was found for MCG-05-23-16 ($<$3.2 per cent). Monte Carlo simulations, conducted to investigate the coronal geometry of these AGN, favor a radially extended corona geometry in NGC~4151 and IC~4329A, a scenario consistent also with MCG-05-23-16, if the disk inclination angle is below 50$^\circ$.
\keywords{Galaxies: active -- Galaxies: groups: individual: NGC~4151, MCG-05-23-16 and IC~4329A  -- polarization }
}
\maketitle{}

\section{Introduction}

The Imaging X-ray Polarimetry Explorer \citep[{\it IXPE};][]{weisskopf22}, is a NASA/ASI mission and the first X-ray imaging polarimeter in orbit in four decades. The three telescopes on board, equipped with polarization-sensitive imaging detectors \citep[Gas Pixel Detector;][]{costa01}, enables pioneering X-ray polarimetric studies on Active Galactic Nuclei (AGN). 

In unobscured radio-quiet (RQ) AGN, the predominant component within the {\it IXPE} energy band (2-8 keV) is the primary X-ray radiation. According to the Unified Model of AGN, this emission is produced by the so-called X-ray corona through inverse Compton scattering of UV photons emitted from the disk \citep{haardt91, haardt93}. This region, composed of extremely hot (with a typical temperature of kT$_{e} \simeq 10-100$ keV; e.g., \citealt{tortosa18}) optically thin electron plasma, is generally placed at a distance of few gravitational radii from the accretion disk. However, the geometry and its precise location have always been debated: from models with large electron heating volumes (e.g., \textit{slab}-corona models; \citealt{haardt91, haardt93, merloni03}) to compact geometrical models (e.g., lamppost geometry; \citealt{martocchia96,fabian17}. The latter possibly originating from a broken jet (\citealt{henri97,ghisellini04}).
Given the scattering nature of the primary X-ray radiation, this emission is expected to present a polarized signal shaped by the structure of the scattering material. Hence, the ongoing investigation of this signal with {\it IXPE} is bringing valuable insights on the symmetry and morphology of the corona. In the first two years of operation, {\it IXPE} observed three unobscured RQ AGN: MCG-05-23-16 studied by \citet{marinucci22} and \citet{tagliacozzo23} (`M22' and `T23' hereafter), NGC~4151 analyzed by \citet{gianolli23} (`G23' hereafter) and IC~4329A studied by \citet{ingram23} (`I23' hereafter). 

In the following, we present the results of the spectro-polarimetric analyses performed on these three AGNs: NGC~4151 in Section \ref{4151}, IC~4329A in Section \ref{ic} and MCG-05-23-16 in Section \ref{mcg}. In Section \ref{conclusion}, we will discuss the common characteristics of these sources.

\begin{figure*}[t!]
\centering
\resizebox{0.8\hsize}{!}{\includegraphics[width=.6\textwidth]{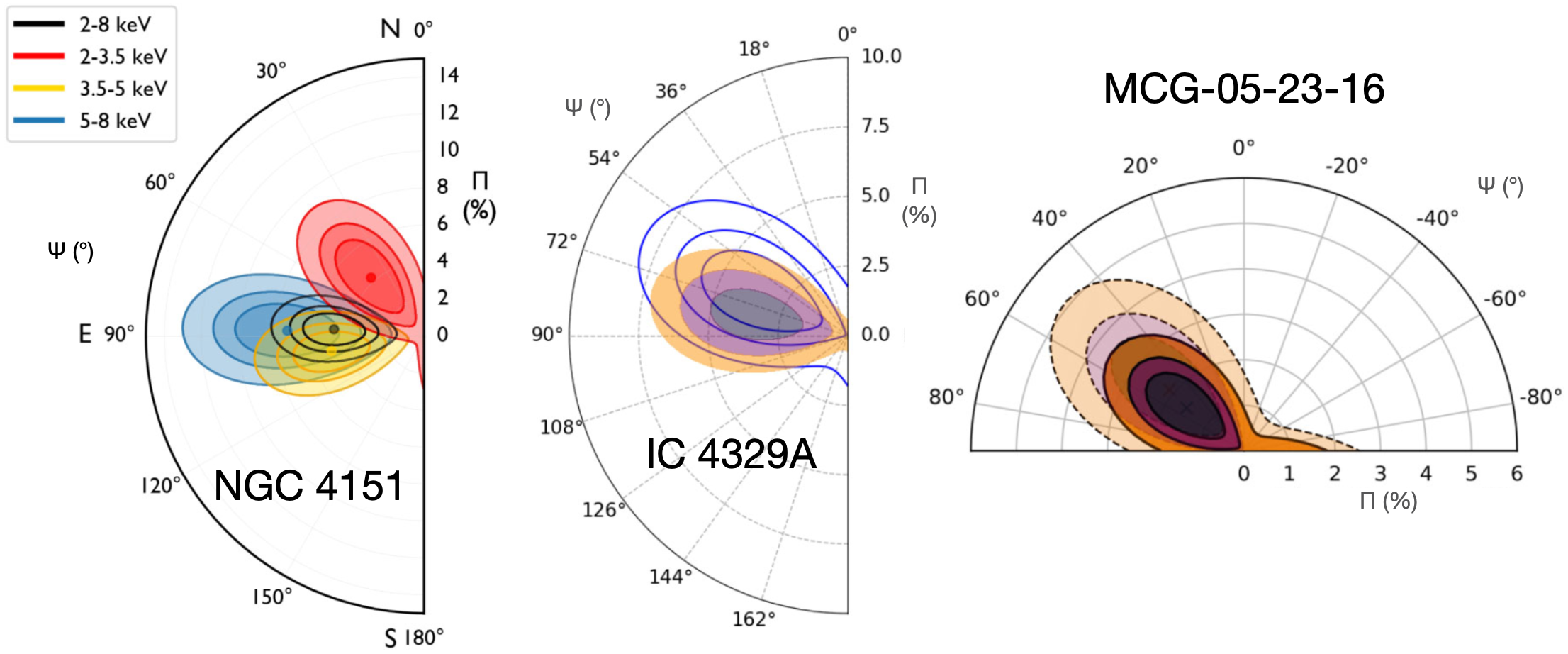}}
\caption{\footnotesize
{ Polarization contours (68, 90, and 99 per cent confidence levels for two degrees of freedom) for the polarization degree $\Pi$ and the polarization angle $\Psi$. {\it Left panel:} polarization properties for different energy ranges in NGC~4151. {\it Central panel:} \texttt{PCUBE} unweighted analysis results (blue) and weighted spectro-polarimetric analysis within XSPEC (solid colors) of IC~4329A. {\it Right panel:} MCG-05-23-06 parametric results for combined (May + November) and first (May) observations. Each contour plot has been taken from the respective original paper.
}}
\label{fig:1}
\end{figure*}

\begin{figure*}[t!]
\centering
\resizebox{0.8\hsize}{!}{\includegraphics[width=\textwidth]{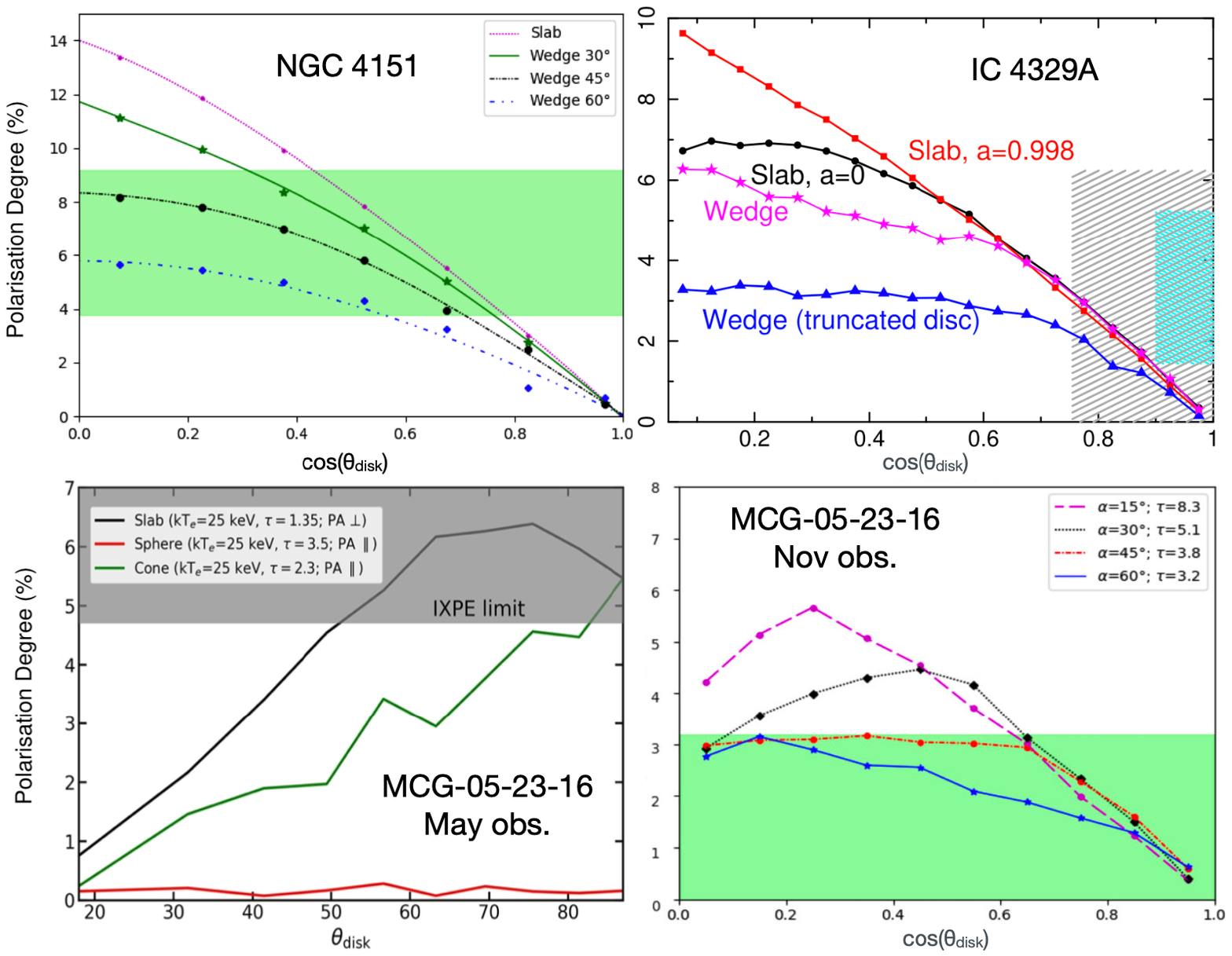}}\\
\caption{\footnotesize
{ Monte Carlo simulations performed with the Comptonization code MONK. {\it Upper panel:} NGC~4151 (left plot) and IC~4329A (right plot). {\it Lower panel:} MCG-05-23-16 May (left plot) and November (right plot) observations; for the second observation, the only tested geometry is the {\it wedge}-like.
The colored bands show the polarization degree ranges obtained from the analyses. Each plot has been taken from the respective original paper.}}
\label{fig:3}
\end{figure*}

\section{NGC~4151}\label{4151}

NGC~4151 is the first RQ unobscured AGN with a constrained X-ray polarization measure for the corona. The source was observed with {\it IXPE} in 2022 December (632 ks), simultaneously to XMM-{\it Newton} (for 33 ks) and {\it NuSTAR} (97 ks).
To assess the polarization properties (i.e., polarization degree PD and polarization angle PA) of the X-ray corona with a model-independent method, G23 adopted the \texttt{PCUBE} analysis, which computes I-normalized Stokes parameters Q and U from selected events. The PD and PA with errors are then derived from Q/I and U/I parameters following the \citet{kislat15} method.
With a detection significance above 99.99\% confidence level ($\sim$4.4$\sigma$), the detected polarization degree is $\Pi$ = 4.9\% $\pm$ 1.1\% and polarization angle $\Psi$ = 86$^\circ$ $\pm$ 7$^\circ$ (east of north) in the {\it IXPE} energy band (see left panel Fig.~\ref{fig:1}).

From the spectral analysis, G23 assessed that the X-ray primary emission is the predominant component in the {\it IXPE} energy band of NGC~4151. Despite its limited spectral contribution reaching up to 6\%, the reflection component has also been detected in the 2-8 keV band. This component arising from the reprocessing of the primary continuum off surrounding material, such as the accretion disk or the torus, presents its own polarization properties \citep{podgorny22}. However, in order to change the derived continuum polarization, the reflection has to be much more strongly polarized than observed by G23. Hence, ensuring that the detected PD and PA truly belong to X-ray primary continuum.
Interestingly, the authors suggest the influence of a further component in 2-3.5 keV band with different polarization properties (PD = 4.3\% $\pm$ 1.6\% and PA = 42$^\circ$ $\pm$ 11$^\circ$). Once taken into account, and by setting the polarization angles of the primary and reflection components to differ by 90 degrees, a PD = 7.7\% $\pm$ 1.5\% and PA = 87$^\circ$ $\pm$ 6$^\circ$ for the X-ray primary emission are found.

Given the high PD measured (i.e., both from the spectro-polarimetric and the \texttt{PCUBE} analysis) and the indication that the polarization is occurring on the equatorial plane (as suggested by the X-ray PA being in the same direction of the radio emission, $\sim$83$^\circ$, \citealp{Harrison1986,ulvestad98}, and reference therein), a ``spherical'' lamppost geometry (with expected PA perpendicular to the disk axis) can be immediately ruled out.

To interpret the obtained polarimetric results, the authors performed simulations adopting the general relativistic Monte Carlo radiative transfer code MONK \citep{zhang19} for coronal geometries with PA consistent to that observed (i.e. parallel to the accretion disk axis) and high PD. The adopted configurations are {\it slab} and {\it wedge}-like geometries (see Fig. \ref{fig:3}). Particularly, in the {\it slab}-like, the corona is above and below the accretion disk, while the {\it wedge} assumes a truncated accretion disk at which end the X-ray corona extends up to the Innermost Stable Circular Orbit (ISCO) with a certain opening angle ($\alpha$, measured from the disk plane), acting as a ``hot accretion flow''. According to these simulations, both geometries reproduce the observed PD and the disk inclination angle derived from X-ray polarization better aligns with that estimated from BLR reverberation studies (i.e. $\sim$58$^\circ$, \citealt{bentz22}). In agreement with a radially extended X-ray corona geometry, the authors observe that the detected X-ray PD is larger than that from UV, optical, and infrared polarization. Meanwhile, the consistent PAs across different spectral ranges implies predominant reprocessing along the equatorial plane.

\section{IC~4329A}\label{ic}

{\it IXPE} observed IC~4329A on 2023 January (for 458 ks), simultaneously to XMM–{\it Newton} (62 ks) and {\it NuSTAR} (for 82 ks).
In order to obtain the polarization properties in the 2–8 keV band, I23 adopted a weighted analysis, fitting I, Q, and U {\it IXPE} spectra with a simple spectro-polarimetric model (see I23 for further details). They found $\Pi$ = 3.3\% $\pm$ 1.1\% and $\Psi$ = 78$^\circ$ $\pm$ 10$^\circ$ (east of north, at 68\% c.l.), corresponding to an upper limit of PD $\leq$ 6.2\% at 99\% c.l. (see central panel Fig.~\ref{fig:1}). Results, verified and consistent with those obtained through a \texttt{PCUBE} analysis, reveal that the obtained X-ray PA (within the errors) closely aligns with the radio emission of the AGN \citep[$\sim$90$^\circ$; see][]{unger87}. Hence, favoring a radially extended corona over a vertically extended one.

The MONK simulations (see Fig. \ref{fig:3}) take into account the following geometries: 1) {\it slab} geometry (extending from 6r$_g$ to 100$_g$) with spin a = 0; 2) {\it slab} extending to r$_{\mathrm{ISCO}}$ for a maximally spinning BH; 3) uniform density {\it wedge}-like geometry with $\alpha$ = 45$^\circ$ and where the disk extends within the corona down to r$_{\mathrm{ISCO}}$; 4) {\it wedge} geometry (with $\alpha$ = 45$^\circ$) and the disk results to be truncated at the outer radius of the corona. 
Even though the two {\it wedge} geometries give a lower PD than the {\it slab}-like, due to their higher symmetry, the observed polarization degree always exceeds those simulated. As explanation, the authors propose that coronal electrons might be outflowing with a mildly relativistic bulk velocity away from the disk plane. Thus, due to relativistic aberration, the predicted polarization would reach the observed value (see I23;\citealt{poutanen23}). 
The same outflowing model would allow the {\it wedge} and truncated disc geometries to replicate our observed polarization properties.

\titlespacing*{\section}
{0pt}{.7ex plus 1ex minus .2ex}{1.ex plus .1ex}
\section{MCG-05-23-16}\label{mcg}

MCG-05-23-16 was observed twice: the first observation took place in 2022 May with XMM–{\it Newton} (58 ks), {\it NuSTAR} (83 ks), and {\it IXPE} (486 ks), while the second was carried out in 2022 November with {\it IXPE} (640 ks) and {\it NuSTAR} (165 ks).
From the polarimetric analysis (see right panel Fig.~\ref{fig:1}) of May and November observations (and their combination), only upper limits to the primary continuum PD were found at 99\% c.l. (first observation: $\leq$ 4.7; second observation: $\leq$ 3.3; and the combination: $\leq$3.2). 

For MCG-05-23-16, MONK simulations were run considering a \textit{slab}-like geometry, a ``spherical'' lamppost on the system symmetry axis, and a truncated cone in outflow (i.e., possible base of a failed jet). Additionally, T23 focused on the {\it wedge}-like geometry. The simulations for the first observation (Fig.~\ref{fig:3}) reveal that the polarization properties of the source can be explained by a ``spherical'' lamppost or a conical geometry of the corona. Alternatively, even if the detected PD is only an upper limit, M22 and T23 show how the PA of the two observations (50$^\circ$ $\pm$ 24$^\circ$, 57$^\circ$ $\pm$ 27$^\circ$ and 53$^\circ$ $\pm$ 13$^\circ$ for the combined data at 68\% c.l.) would align with the Narrow Line Region (NLR) of the AGN \citep[$\sim$40$^\circ$][]{ferruit00}. Thus, assuming an accretion disk perpendicular to the NLR, a \textit{slab}-like geometry can also be taken into account for disk inclination angles $<$50$^\circ$ (see M22).
As for the {\it wedge} geometry, given the unconstrained disk inclination, a broad range of opening angles can be considered to explain the polarimetric results (see T23 for more details and Fig.~\ref{fig:3}). In summary, all the tested geometries potentially reproduce the observed upper limit in PD, and the absence of an independent constraint on source inclination hinders definitive conclusion. However, assuming the disk inclination recently measured by \citet{serafinelli23}, $\sim$41$^{+9}_{-10}$$^\circ$, a {\it wedge}-like geometry is favored over a conical corona.

\section{Conclusions}\label{conclusion}

From the polarimetric studies conducted on the three unobscured RQ AGN, we observe that the polarization degrees are broadly consistent within each other at 1$\sigma$ c.l.: NGC~4151 with 4.9\% $\pm$ 1.1\%, IC~4329A with 3.3\% $\pm$ 1.1\% and the upper limit of MCG-05-23-16 $<$3.2\%. The found polarization angles well align with the orientation of the radio emission or the NLR, suggesting that the preferred coronal geometry is a radially extended {\it slab}-like geometry. However, the uncertainties on the disk inclinations is preventing from distinguishing between analogous geometries, such as between {\it slab}-like and {\it wedge}-like configurations in the cases of NGC~4151 and IC~4329A.

These initial discoveries represent a fundamental advancement in comprehending the X-ray corona geometry in AGN and the forthcoming {\it IXPE} observations hold the promise of substantially impact our current knowledge in this field. \\

\footnotesize
\noindent {\bf{Affiliations}}\par
$^{6}$Dipartimento di Fisica, Università degli Studi di Roma ‘La Sapienza’, Piazzale Aldo Moro 5, I-00185 Roma, Italy\par
$^{7}$Dipartimento di Fisica, Università degli Studi di Roma ‘Tor Vergata’, Via della Ricerca Scientifica 1, I-00133 Roma, Italy\par
$^{8}$Université de Strasbourg, CNRS, Observatoire Astronomique de Strasbourg, UMR 7550, F-67000 Strasbourg, France\\
$^{9}$Istituto Nazionale di Fisica Nucleare, Sezione di Roma ‘Tor Vergata’, Via della Ricerca Scientifica 1, I-00133 Roma, Italy \\
$^{10}$Department of Astronomy, University of Maryland, College Park, MD 20742, USA 
\begin{acknowledgements}
The Imaging X-ray Polarimetry Explorer ({\it IXPE}) is a joint US and Italian mission. The US contribution is supported by the National Aeronautics and Space Administration (NASA) and led and managed by its Marshall Space Flight Center (MSFC), with industry partner Ball Aerospace (contract NNM15AA18C). The Italian contribution is supported by the Italian Space Agency (Agenzia Spaziale Italiana, ASI) through contract ASI-OHBI-2017-12-I.0, agreements ASI-INAF-2017-12-H0 and ASI-INFN-2017.13-H0, and its Space Science Data Center (SSDC) with agreements ASI- INAF-2022-14-HH.0 and ASI-INFN 2021-43-HH.0, and by the Istituto Nazionale di Astrofisica (INAF) and the Istituto Nazionale di Fisica Nucleare (INFN) in Italy. This research used data products provided by the IXPE Team (MSFC, SSDC, INAF, and INFN) and distributed with additional software tools by the High-Energy Astrophysics Science Archive Research Center (HEASARC), at NASA Goddard Space Flight Center (GSFC).

\end{acknowledgements}

\bibliographystyle{aa}
\bibliography{biblio}

\end{document}